\documentclass[%
 aip,
 amsmath,amssymb,
 reprint,%
]{revtex4-1}
\usepackage{graphicx}% Include figure files
\usepackage{dcolumn}% Align table columns on decimal point
\usepackage{bm}% bold math
\usepackage[utf8]{inputenc}
\usepackage[T1]{fontenc}
\usepackage{mathptmx}
\usepackage{etoolbox}

\makeatletter
\def\@email#1#2{%
 \endgroup
 \patchcmd{\titleblock@produce}
  {\frontmatter@RRAPformat}
  {\frontmatter@RRAPformat{\produce@RRAP{*#1\href{mailto:#2}{#2}}}\frontmatter@RRAPformat}
  {}{}
}%
\makeatother
\begin{document}

\preprint{AIP/123-QED}

\title{Optimizing Low-Energy Anti-Fibrillation Pacing:\\Lessons from a Cellular Automaton Model}
% Force line breaks with \\
\author{N. DeTal}
 \email{nddetal@gmail.com.}
\author{F. Fenton}
\affiliation{ 
School of Physics, Georgia Institute of Technology, Atlanta, Georgia 30332, USA%\\This line break forced with \textbackslash\textbackslash
}%

\date{\today}% It is always \today, today,

\begin{abstract}
The essential features of far-field low-energy defibrillation are elucidated using a simple cellular automaton model of excitable media. The model's topological character allows for direct correspondence with both realistic models and experiment. An optimal pacing period is shown to arise from the competition between two effects, and not a resonant response as was previously hypothesized. Finally, a topologically motivated feedback scheme is presented that outperforms traditional LEAP by identifying optimal shock timings.
\end{abstract}

\maketitle

% \begin{quotation}
% The ``lead paragraph'' is encapsulated with the \LaTeX\ 
% \verb+quotation+ environment and is formatted as a single paragraph before the first section heading. 
% (The \verb+quotation+ environment reverts to its usual meaning after the first sectioning command.) 
% Note that numbered references are allowed in the lead paragraph.
% %
% The lead paragraph will only be found in an article being prepared for the journal \textit{Chaos}.
% \end{quotation}

\section{\label{sec:intro}Introduction}

Cardiac fibrillation is often driven by reentrant spiral waves of electrical activity in the heart \cite{davidenko1992stationary,ricknat,cherry2008visualization}. These patterns are defined by topological singularities near their center of rotation \cite{winfree,winfreetop,experimentalist}. While traditional defibrillation applies a single strong shock to remove all singularities and reset the tissue \cite{fenton2009termination}, recent theory \cite{weh,markusbar} and experiment \cite{efimov,leap,leapsync} has shown that the same result can be achieved with a series of weaker shocks. This therapy is known as low-energy anti-fibrillation pacing (LEAP). Extensive simulations have been performed analyzing LEAP using reaction-diffusion models in a heterogeneous domain \cite{leapsync,markusbar}. However, the complexity of such detailed models obscures the fundamental mechanisms responsible for successful defibrillation. In this paper, we show how significant insight can be gained from simulations in a very simple cellular automaton model. The model's minimal structure allows for easy tracking of individual spiral core singularities even in the presence of heterogeneous excitation. As a result, we are able to clearly demonstrate through simulation why certain pacing periods lead to more effective defibrillation. In spite of its simplicity, the model crucially retains the topological structure of more realistic models. By clarifying the role of topological constraints on the defibrillation process, we ultimately deduce a further optimized pacing strategy using a model-independent framework.

\section{\label{sec:ca}Cellular Automaton Model of LEAP}

The simplest possible model of excitable media is the Greenberg-Hastings (GH) cellular automaton \cite{ghca}. In this model, each lattice cell may occupy one of three states: resting, excited, or refractory. At each discrete time step, excited cells become refractory, refractory cells become resting, and resting cells become excited if a neighboring cell is excited. In two spatial dimensions, these rules are sufficient to support both unidirectional traveling waves and rotating spiral waves. In their original study, Greenberg and Hastings identified a conserved winding number that can be defined at each vertex on the lattice \cite{combinatorics}. Spiral wave cores have a winding number of $\pm 1$ (with chirality given by the sign) and thus persist indefinitely, generating chaotic fibrillating states. An important consequence is that core singularities, and hence the spirals they belong to, must be eliminated for fibrillation to cease. This feature is analogous to the phase singularities of spiral waves in continuous systems \cite{winfree,winfreetop}.

\begin{figure}
    \centering
    \includegraphics[scale=.35]{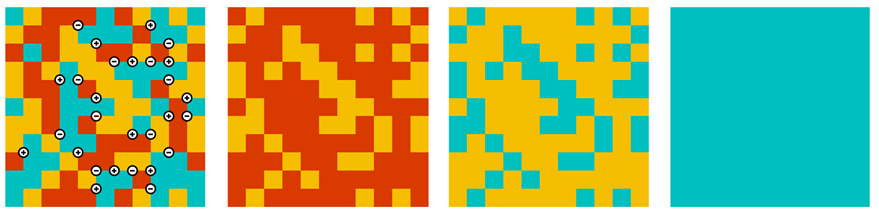}
    \caption{Complete defibrillation of a complex state containing many spiral singularities. Every resting cell is excited, resulting in immediate annihilation of every singularity. Singularities are identified by nonzero winding number and are marked with a $\pm$ corresponding to chirality. Excited cells are red, refractory cells are yellow, and resting cells are turquoise}
    \label{fig:defib}
\end{figure}

In order to gain insight into the defibrillation process, we study the GH model in the presence of stimulation (or shocking) which we implement by spontaneously making resting cells excited. Traditional defibrillation uses a single large shock to excite and reset all tissue in the heart simultaneously; while initially unexplained, it is now understood that this method succeeds by removing all reentrant spiral wave singularities \cite{fenton2009termination}. This feature is retained in the modified GH model, as shown in Figure \ref{fig:defib}. In all figures, excited cells are red, refractory cells are yellow, and resting cells are turquoise. If every resting cell is excited, the winding number at every vertex must be zero, and no spirals can remain. This can be understood intuitively by considering the cores as phase singularities where all possible values of phase converge. By making all resting cells excited, the resting ``phase'' will necessarily be missing and singularities can no longer be present.

While theoretically a strong shock that excites the entire domain is always successful at removing spiral waves, clinical defibrillation is painful and can damage heart tissue in the process \cite{ouch}. There is therefore a strong clinical interest in developing effective low-energy defibrillation strategies. LEAP uses a series of multiple weak shocks to achieve the same results as traditional defibrillation \cite{leap, leapsync, markusbar}. Both methods exploit the virtual electrode phenomenon, in which far-field stimulation generates localized tissue activation around the heart's many natural heterogeneities \cite{ve1,ve2}. Stronger stimulation can recruit ever smaller heterogeneities, exciting a larger fraction of the total domain. Tissue response is therefore governed by size and density distributions of the heterogeneities as well as the stimulus strength. We characterize the tissue response for the cellular automaton model with a single parameter $p$---the probability for a resting cell to become excited during a shock. When $p=1$, all resting cells are excited and complete defibrillation is recovered. For $p<1$, a single shock, in general, will no longer completely defibrillate. We refer to $p$ as the shock strength, but it encapsulates both external stimulation strength and heterogeneity density in a simplified fashion.

\begin{figure}
    \centering
    \includegraphics[scale=.25]{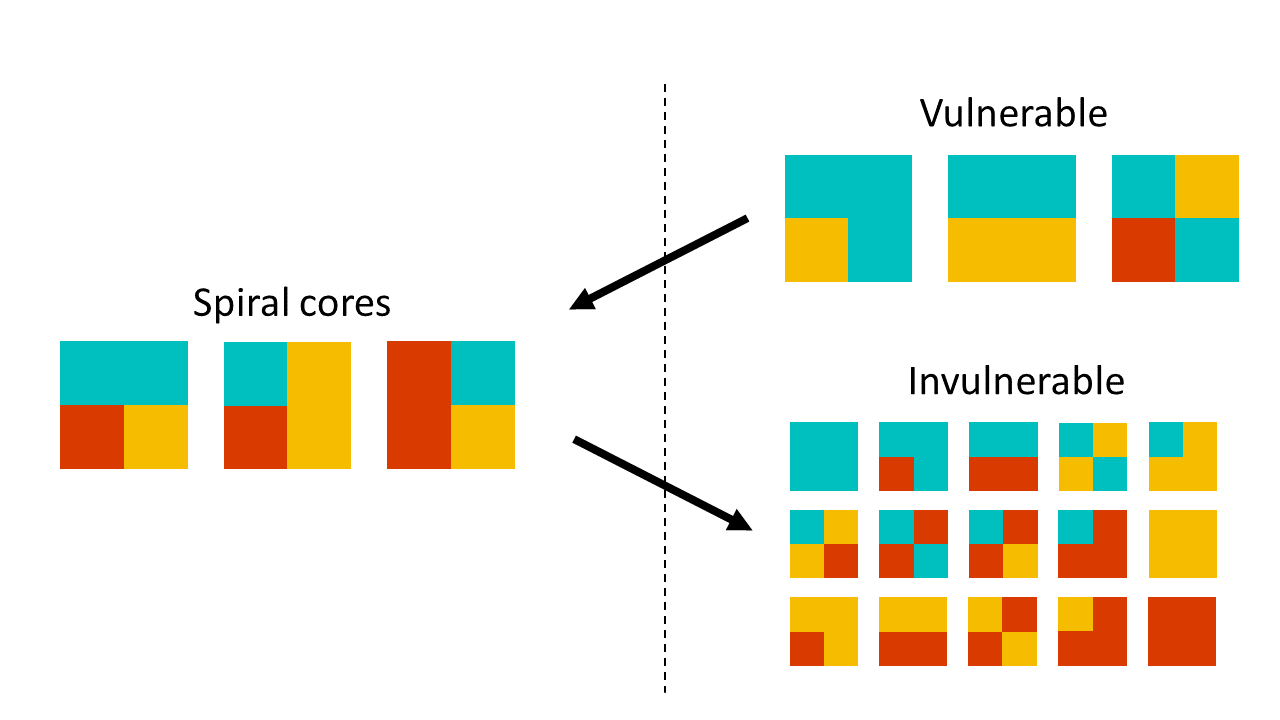}
    \caption{Categorization of all possible $2\times2$ block patterns up to reflections and rotations. Spiral cores (C) have nonzero winding number and generate persistent spiral waves. Vulnerable patterns (V) may be converted to cores by a shock of strength $p<1$ while invulnerable patterns (I) are immune to such a conversion. Cores can be converted to invulnerable patterns yielding partial defibrillation.}
    \label{fig:cvi}
\end{figure}

For the original GH model, every $2\times2$ block of cell configurations can be categorized by the topological winding number at its vertex. This results in two distinct categories: spiral core patterns and non-core patterns. With the inclusion of the shocking procedure, non-core patterns can be further categorized as either vulnerable or invulnerable patterns. During a shock step, vulnerable patterns can be converted into core patterns, while invulnerable patterns cannot. Core patterns can also be converted to invulnerable patterns, in which case the core's spiral is eliminated. Figure \ref{fig:cvi} shows this categorization for all unique $2\times2$ block patterns. Low-energy shocks can thus fail to defibrillate not only because they do not remove all existing spiral core singularities, but also because they may generate additional singularities. This effect has been documented in experiments on isolated rabbit and dog hearts \cite{ve2,efimovinduced} and is closely connected to the S1-S2 initiation of spiral waves commonly demonstrated in computational models \cite{s1s22,s1s2}. Remarkably, members of the three categories share identical transition probabilities to the other categories given by the following transition matrix:

\begin{equation}
    \begin{pmatrix}
    C_{t+1}\\
    V_{t+1}\\
    I_{t+1}
    \end{pmatrix}=\begin{pmatrix}
    1-p & 2p(1-p) & 0\\
    0 & (1-p)^2 & 0\\
    p & p^2 & 1
    \end{pmatrix}
    \begin{pmatrix}
    C_{t}\\
    V_{t}\\
    I_{t}
    \end{pmatrix}
    \label{eq:transition}
\end{equation}
In Section \ref{sec:tophat}, we discuss how this structure is related to the underlying topology of  singularities, and how it can be exploited to optimize low-energy pacing.

\begin{figure}
    \centering
    \includegraphics[scale=.35]{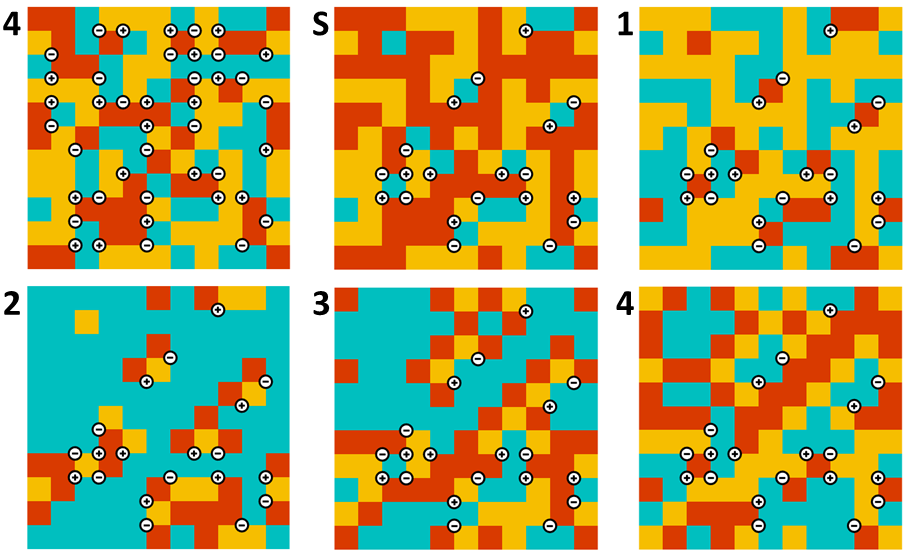}
    \caption{A full period of evolution for a LEAP simulation with $N=10$, $T=4$, $p=0.7$. The shock step S reduces the total number of cores.}
    \label{fig:leapex}
\end{figure}

LEAP is typically implemented by weak pacing at regularly timed intervals. We mimic this in our cellular automaton simulations as follows. First, we randomly initialize the cells on a $N\times N$ domain, resulting in a complex state containing many spiral cores. We then apply a period $T$ of ordinary time-evolution steps followed by a single shock step with strength $p$. This sequence of $T+1$ steps is then repeated until the total number of cores reaches a statistical steady state in which, on average, the number of cores generated by a shock balances the number of cores destroyed. To ensure that the total topological winding number is zero and cores exist in pairs of opposite chirality, we employ periodic boundary conditions. Figure \ref{fig:leapex} shows the beginning of a LEAP simulation with $N=10$, $T=4$, $p=0.7$. 

\begin{figure}
    \centering
    \includegraphics[scale=.4]{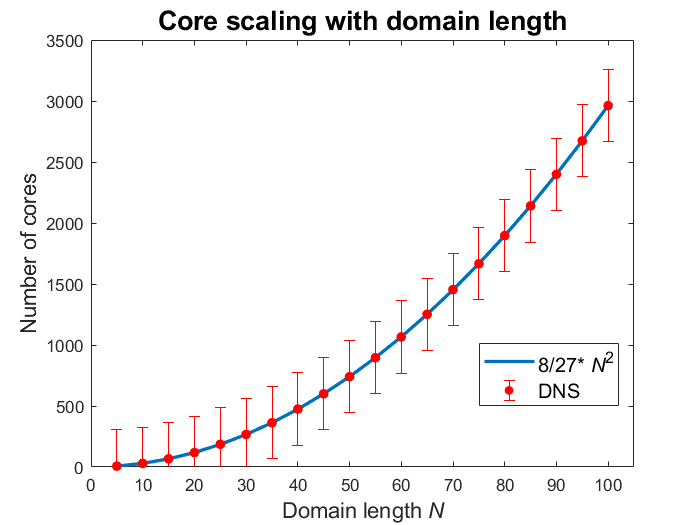}
    \caption{Number of cores from random initial conditions as domain size is varied. Statistical results show scaling proportional to the total domain area. Error bars are scaled by $1/N^2$ and also scale linearly with area.}
    \label{fig:cores}
\end{figure}

\begin{figure}
    \centering
    \includegraphics[scale=.4]{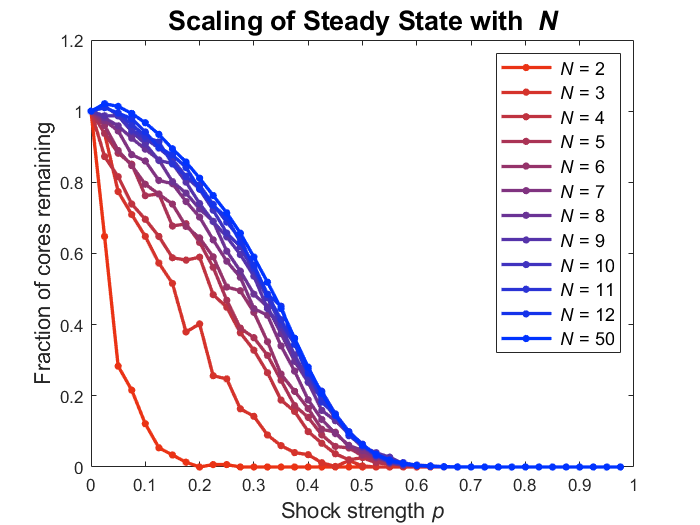}
    \caption{Steady state number of cores scaled by relation (\ref{scaling}) in LEAP simulation for $T=4$. Results converge near $N\approx 50$.}
    \label{fig:leaps}
\end{figure}

For a $2\times 2$ domain, the average number of cores $\overline W$ generated from random initial conditions can be easily calculated by enumeration to be $\frac{32}{27}$. The statistical results from Figure \ref{fig:cores} show that this result extrapolates to larger domains by scaling with the total area (or alternatively, number of vertices) such that the general result is
\begin{equation}
    \overline{W}(N) = \frac{8}{27}N^2.
    \label{scaling}
\end{equation}
This provides a convenient extensive scaling for comparing results between domains of different sizes. In particular, we are interested in the statistical steady state as $N\rightarrow \infty$. Figure \ref{fig:leaps} shows the LEAP steady state number of cores scaled by (\ref{scaling}) for $T=4$ as $p$ and $N$ are varied. By $N=50$, steady state results already converge. For subsequent LEAP simulation statistics, we thus take $N=50$ and normalize the number of cores by (\ref{scaling}) to obtain intensive results absent of finite-size effects.

\begin{figure}
    \centering
    \includegraphics[scale=.4]{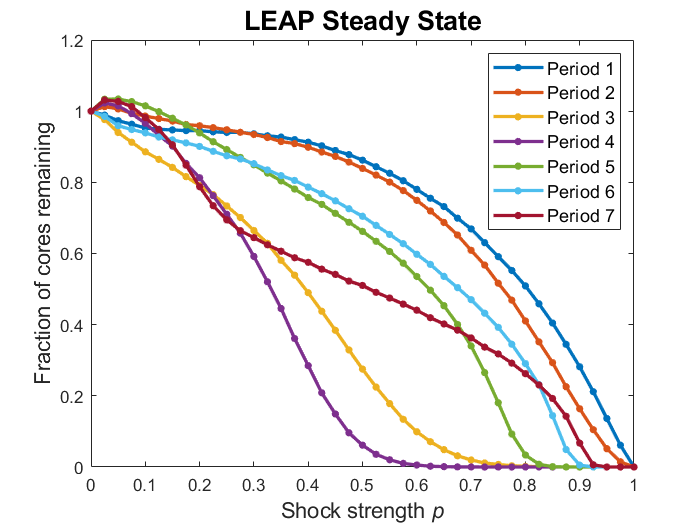}
    \caption{Steady state number of cores after many shocks in LEAP simulations as period and shock strength are varied.}
    \label{fig:leapresults}
\end{figure}

\begin{figure}
    \centering
    \includegraphics[scale=.4]{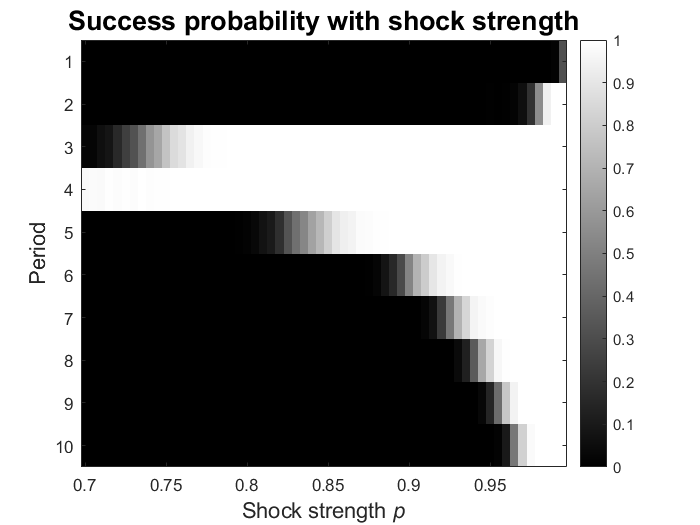}
    \caption{LEAP success probability after 20 shocks as period and shock strength are varied. A strong peak is localized near the $T=4$ spiral rotation period.}
    \label{fig:sux}
\end{figure}

Figure \ref{fig:leapresults} shows the steady state results of LEAP simulations as period and shock strength are varied. While the average number of cores decreases with increased shock strength for all periods, certain periods perform particularly well. Period 4---the period of rotation for individual spirals---in particular is able to defibrillate in finite time when as little as 60\% of tissue is excited per shock. Figure \ref{fig:sux} shows the probability of successful defibrillation after 20 shocks and highlights the peak in efficiency near period 4.

\section{\label{sec:origin}Origin of an optimal pacing period}
A commonly reported feature of LEAP in both simulation \cite{leapsync,markusbar} and experiment \cite{leap,leapsync,fenton2009termination} is the presence of an optimal pacing period close to either the dominant period of fibrillation or, equivalently, the period of spiral rotation. As was demonstrated in Section \ref{sec:ca}, this feature is reproduced even in the highly simplified GH model which suggests that the mechanism responsible is a generic feature of all excitable systems. Previously no explanation for this apparent resonance has been provided. Using the GH model, we now show through simple examples how it arises as a competition between two mechanisms intrinsic to generic excitable media. 

\begin{figure}
    \centering
    \includegraphics[scale=.35]{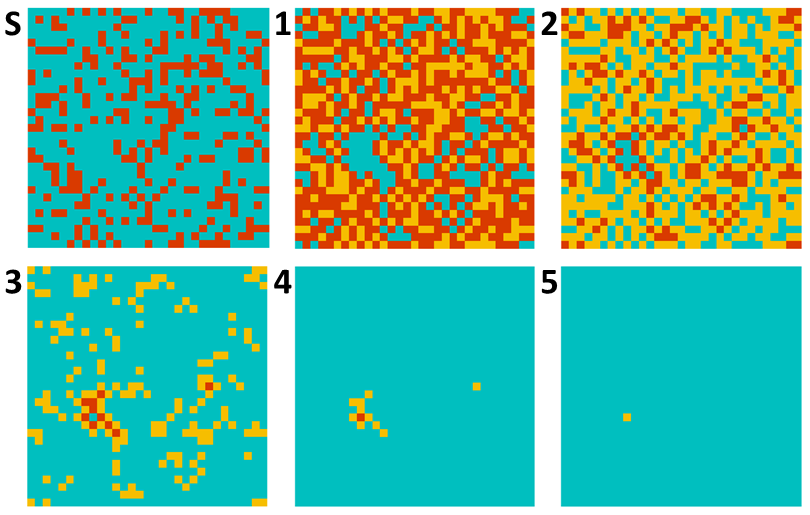}
    \caption{LEAP pacing with $T=7$, $p=.3$ starting from the uniform rest state. The entire domain returns to rest before the next shock is applied and no new cores can be created.}
    \label{fig:longweak}
\end{figure}

From Figure \ref{fig:leapresults}, we see that pacing with $T=7$, $p=.3$ is highly ineffective, reducing the number of cores to only 60\% of the starting number. An entirely different result is achieved when the system is initialized in the the uniform rest state, however. Figure \ref{fig:longweak} shows how the shock is unable to generate any new cores due to a lack of refractory cells. By $t=6$, the system has returned to the uniform rest state a full time step before the next shock is applied. As a result, the process repeats and no new cores are created.

\begin{figure}
    \centering
    \includegraphics[scale=.35]{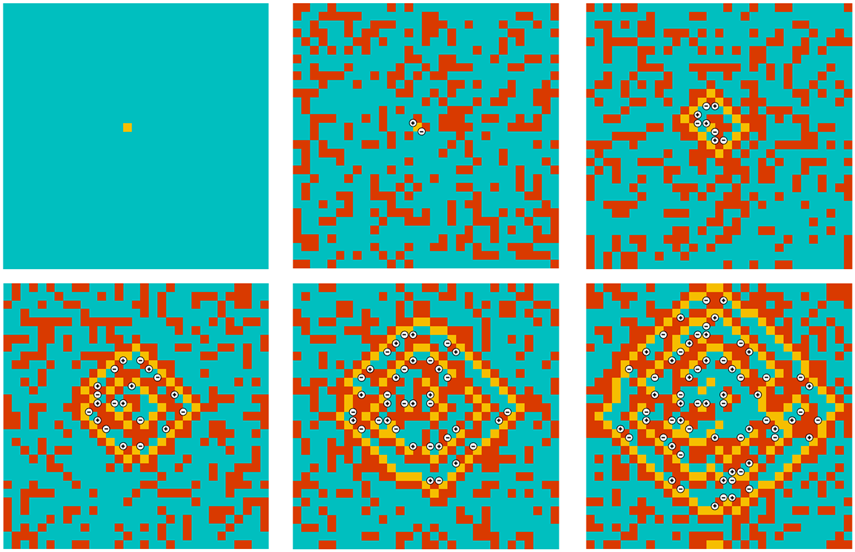}
    \caption{LEAP pacing with $T=7$, $p=.3$ starting from the uniform rest state with a single perturbing refractory cell. Each shock generates new cores nucleating from the initial perturbation. Seven time steps and a single shock step elapse between each plotted frame after the first.}
    \label{fig:longweakpert}
\end{figure}

Because the system is ergodic, it will eventually converge to the steady state reported in Figure \ref{fig:leapresults}, although this metastable sequence is extremely long-lived. Figure \ref{fig:longweakpert} shows how the presence of even a single refractory cell can rapidly trigger a transition to the steady state. The first shock initiates a small number of cores localized to the perturbation. During the time preceding the next shock, the cores generate expanding target waves which contain many additional vulnerable patterns. The next shock thus creates even more cores along the back of this expanding wave. This process repeats, resulting in a nucleation of new cores surrounding the initial perturbation.

By allowing the excitation of the previous shock to fully dissipate before the next shock is applied, longer-period pacing prevents the creation of new cores associated with transient refractory regions. When cores are present, however, the associated vulnerable regions persist and grow as subsequent shocks generate new cores.

\begin{figure}
    \centering
    \includegraphics[scale=.3]{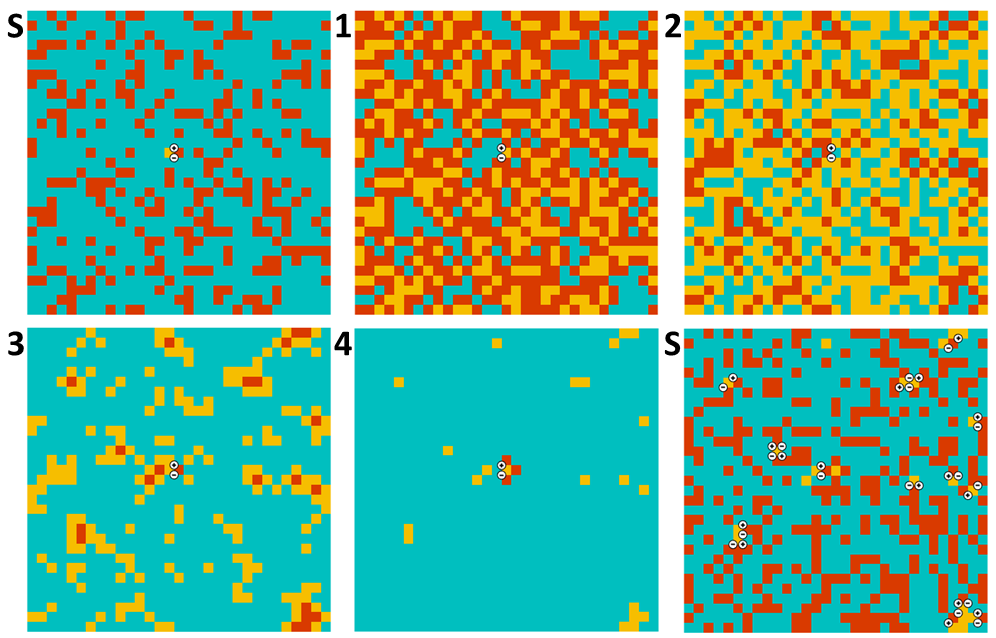}
    \caption{LEAP pacing with $T=4$, $p=.3$ starting from the uniform rest state with a single perturbing refractory cell. No nucleation is observed but many new cores are generated near unrecovered tissue.}
    \label{fig:shortweakpert}
\end{figure}

Figure \ref{fig:shortweakpert} shows the results of pacing the perturbed state with a shorter period of $T=4$. The first shock once again initiates new cores near the perturbing refractory cell. By the time the next shock is applied, the original cores have barely begun to form target patterns and do not produce extra vulnerable regions. However, a number of transient refractory cells still remain from the previous shock. The associated vulnerable patterns of these transients trigger the generation of many new cores uniformly throughout the domain.

\begin{figure}
    \centering
    \includegraphics[scale=.45]{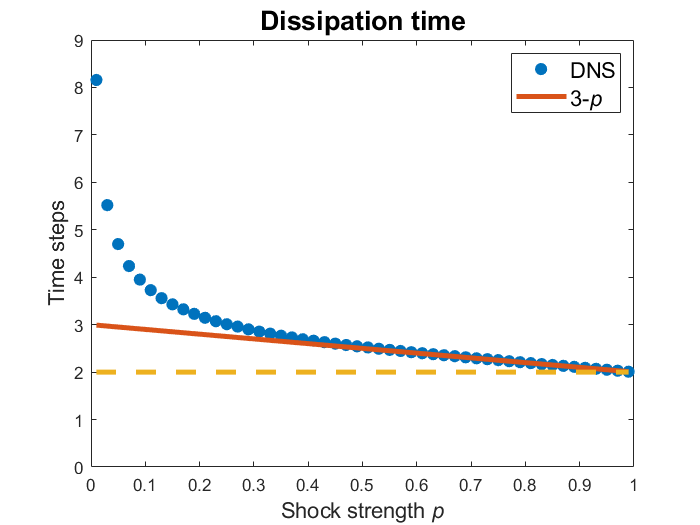}
    \caption{Average amount of time for cells to return to rest after a shock is applied. }
    \label{fig:dissipation}
\end{figure}

\begin{figure}
    \centering
    \includegraphics[scale=.3]{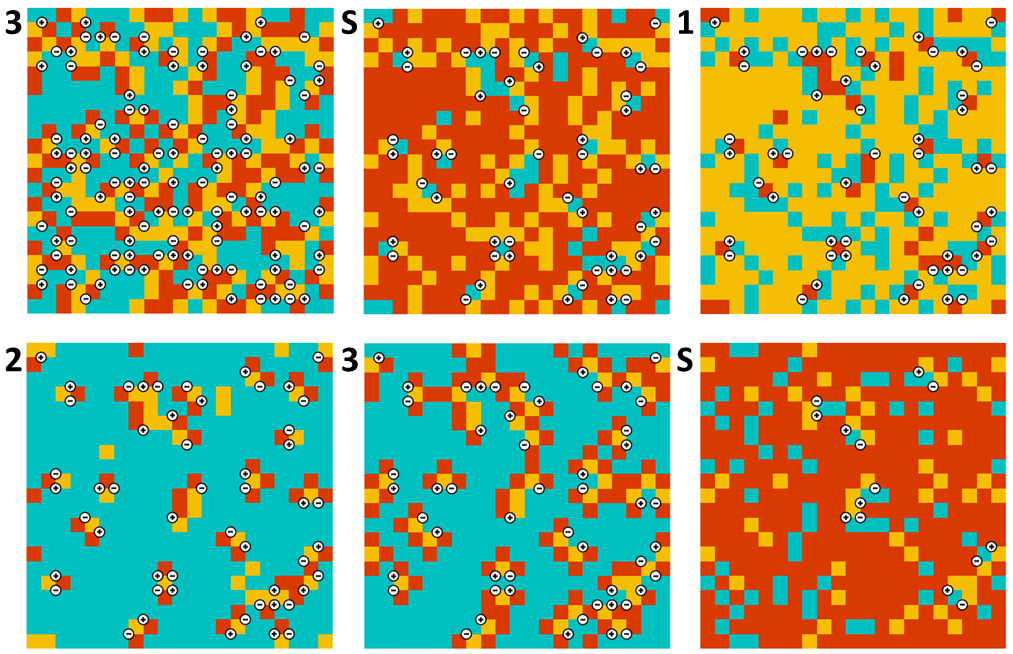}
    \caption{LEAP pacing with $T=3$, $p=.8$. Cores are efficiently eliminated with minimal reinitiation.}
    \label{fig:shortstrong}
\end{figure}

These two examples demonstrate how pacing periods that are either too long or too short are both susceptible to generating new spiral cores and thus ineffective at defibrillating. The optimal period must be short enough to suppress localized nucleation around existing cores but long enough to allow transient refractory cells to recover. A useful quantity to consider in light of these mechanisms is the mean dissipation time. This is the average amount of time for cells to return to rest after a shock is applied. For shocks of moderate strength, rest cells are either shocked with probability $p$ and take two time steps to recover, or else are not shocked but are immediately excited by a neighbor at the next time step and thus require three time steps to recover. The resulting mean dissipation time is given by
\begin{equation}
    \tau \approx 2p+3(1-p) = 3-p.
\end{equation}
Figure \ref{fig:dissipation} shows that this rough estimate is in excellent agreement with direct numerical simulation by $p\approx .5$. As shock strength approaches 1, the mean dissipation time is limited by the two steps a single cell takes to recover. Thus, for moderate shock strength, $T=3$ is the fastest pacing period which avoids core creations due to the transients of previous shocks. Figure \ref{fig:shortstrong} shows the extremely efficient progression of $T=3$ pacing for $p=.8$. After only a few shocks, all cores are removed. 

Rather than a resonance phenomenon, we see that the optimal pacing period emerges from a competition between spiral creation mechanisms occurring at short and long periods. Nevertheless, the optimal period and the spiral period do appear correlated. One possible explanation is that the optimal period and spiral period are both roughly set by the mean dissipation time; in order for a spiral wave to undergo a full rotation, the surrounding tissue must complete a full cycle of recovery. A similar condition is required for efficient pacing. 

If there is no true resonant response at the spiral period, as our analysis suggests, non-periodic pacing has the prospect of providing more efficient defibrillation by allowing for more sophisticated shocking protocols. In the following section we explore this possibility and indeed find that allowing the period between shocks to vary leads to improved efficiency.

%Topologically Optimized Pacing of Heart Arrhythmia Turbulence (TOPHAT)
\section{\label{sec:tophat}Topologically optimized pacing}
Our analysis thus far suggests that optimal low-energy pacing is achieved not by matching the spiral period, but by pacing at times which minimize the probability of creating new spiral wave singularities. Although fixed-timing pacing is optimized when the period approximately equals the mean dissipation time, in general the time for tissue to relax after a shock varies. Moreover, the probability of creating new singularities at a given time depends on the detailed spatial configuration and evolves irregularly due to the chaotic nature of fibrillation. A truly optimal pacing strategy must constantly adapt its timing accordingly. In this section, we will show how topological analysis can be used to determine when an applied shock will be most effective. To this end we first describe the general features of topological analysis in continuous excitable media and how it may be extended to the discrete cellular automaton model.

The topology of spiral wave dynamics has often been characterized by a local phase variable, hence the terminology of ``phase singularity'' \cite{winfree,ricknat}. However, modern work on excitable systems has shown that a more natural topological description is obtained by tracking the regions of tissue that are excited or refractory, respectively \cite{keener,topological,robust,robustexp,ME}. In two-dimensional systems, these regions are bounded by one-dimensional contours that are the primary objects of interest. As a result of continuity, each contour either forms a closed loop or terminates at a boundary \cite{winfreetop}. Singularities are located at the isolated points where the excited and refractory contours intersect. 

\begin{figure}
    \centering
    \includegraphics[scale=.4]{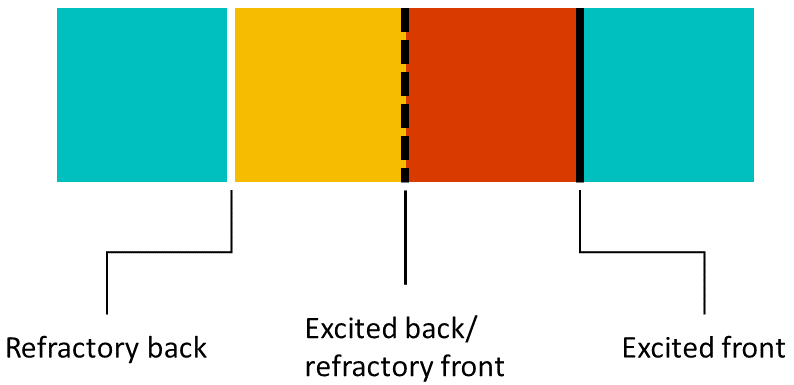}
    \caption{Organizing topological contours (white, dashed black, and black) defined along edges between neighboring cells. }
    \label{fig:contoursdef}
\end{figure}

The excited and refractory contours can be further subdivided into back and front segments to capture the propagating nature of the system dynamics. Excited fronts and refractory backs are formed where the respective contours enter regions in the rest state. Excited backs and refractory fronts are formed in the opposite cases, leading to a total of four segments types \cite{topological}. Propagating waves lead with an excited front that spreads into resting tissue, and is followed (in order) by a refractory front, an excited back, and finally a refractory back that connects back to the rest state. Figure \ref{fig:contoursdef} shows how the contour segments can be defined in the GH model along cell edges. Since excited and refractory regions cannot overlap in this discrete model, the excited back and refractory front lie along the same edge. 

\begin{figure}
    \centering
    \includegraphics[scale=.3]{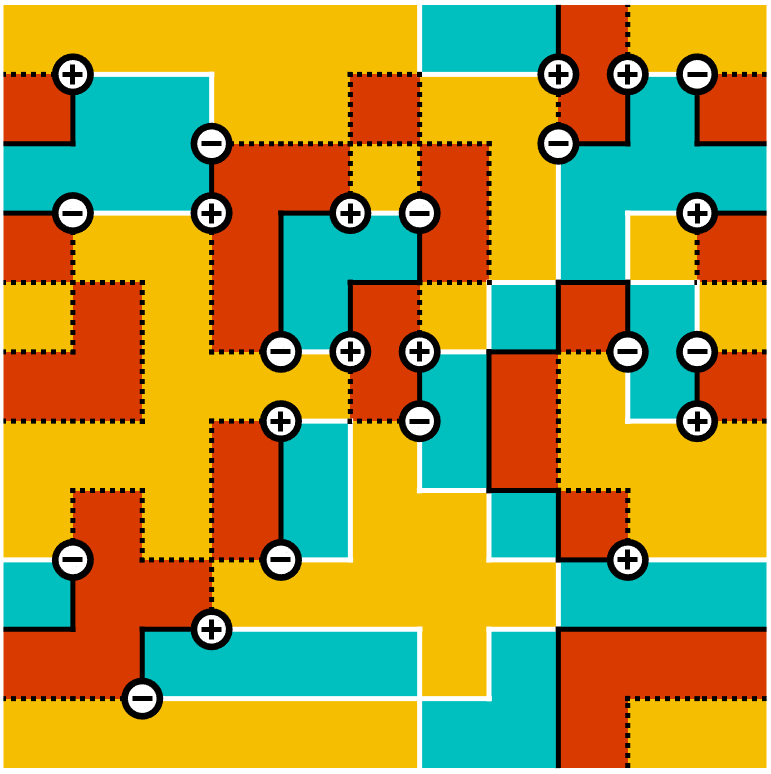}
    \caption{Topological contours and singularities plotted for an arbitrary fibrillating state. White lines are refractory backs, black lines are excited fronts, and dashed black lines are excited backs/refractory fronts.}
    \label{fig:contoursfib}
\end{figure}

\begin{figure}
    \centering
    \includegraphics[scale=.35]{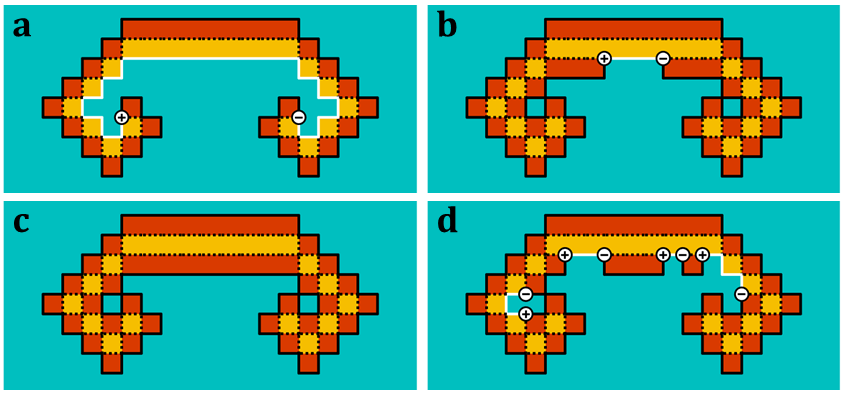}
    \caption{Elimination of singularities via contour stimulation. (a) Initial pair of oppositely rotating singularities joined by all three topological contours. (b) Both singularities are transported closer together by stimulating continuously along the white refractory back contour. (c) Stimulating the entire refractory back contour causes the two singularities to mutually annihilate. (d) Discontinuous stimulus of the refractory back contour from a shock with $p<1$ leads to the creation of several new pairs of singularities.}
    \label{fig:teleport}
\end{figure}

Figure \ref{fig:contoursfib} shows the contours and singularities for an arbitrary fibrillating state and demonstrates a significant topological feature: each singularity is connected by contour segment to a singularity of opposite chirality. This is actually a manifestation of conservation of topological charge \cite{winfreetop,topological,keener}, which states that spiral core singularities can only be created or destroyed in oppositely rotating pairs. 
In particular, it was recently shown \cite{ME} that the necessary condition for eliminating a pair of singularities is to excite along the refractory back contour joining them. This instantaneously removes the two singularities and converts the reentrant spiral waves into a transient radially propagating wave. If only part of the joining refractory back contour is excited, the singularities will in general be transported along it but will not be eliminated. Sufficiently irregular excitation leads to the creation of entirely new pairs of singularities via the S1-S2 creation mechanism \cite{s1s2}. Each of these cases is demonstrated in Figure \ref{fig:teleport} for a single pair of connected singularities in the GH model. Successful single-shock defibrillation thus requires the excitation of every rest cell adjacent to a refractory back contour in order to annihilate every pair of singularities. When all rest cells are excited, as in Figure \ref{fig:defib}, this condition is automatically met without a detailed understanding of the underlying topology. However, the same result can clearly be achieved by shocking a smaller number of cells. 

\begin{figure}
    \centering
    \includegraphics[scale=.4]{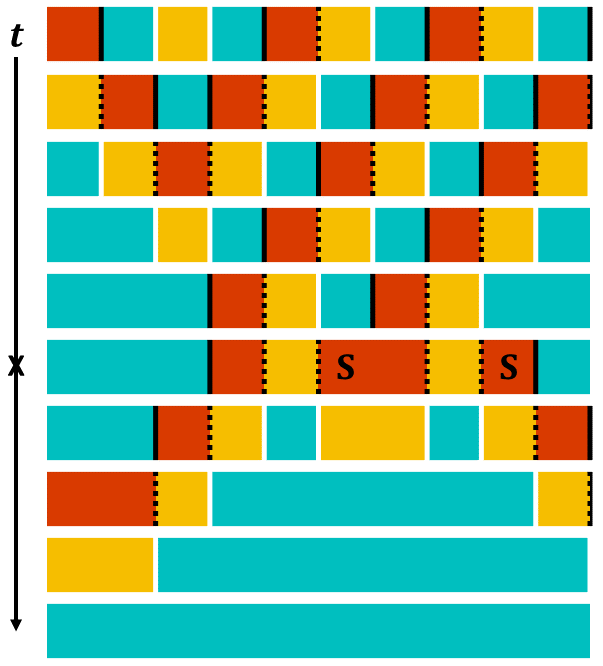}
    \caption{Defibrillation in a one-dimensional periodic cable. The initial configuration contains two leftward-traveling waves and a bidirectional wave representing reentrant fibrillation. At the marked time step (cross), refractory back contours are stimulated leading to defibrillation.}
    \label{fig:cable}
\end{figure}

Excitation of the refractory back contours is also the necessary condition for defibrillation in one-dimensional cables. In this geometry, fibrillation is represented by reentrant traveling waves and can be topologically characterized with a global winding number \cite{roman1d, keener}. Figure \ref{fig:cable} shows the evolution and subsequent defibrillation of a configuration with initial winding number $-2$ corresponding to two leftward-traveling waves and a bidirectional wave with zero net winding number. At the fifth time step, the refractory back contours are stimulated and generate two rightward-traveling waves which then collide and annihilate with the original waves. 

While the topological contour framework clearly dictates the necessary conditions for defibrillation and explains the success of domain-wide stimulation, it also explains why low-energy multi-shock defibrillation can fail. If stimulus along the refractory back contour is not uniform, new pairs of singularities are created as demonstrated in Figure \ref{fig:teleport}(d). In this example, the low-energy shock generated new singularities without removing any of the originals, leading to a worse state of fibrillation. The vulnerable patterns described in Section \ref{sec:ca} are a manifestation of this property; each contains refractory back contours that allow for the stimulated creation of new singularities.

\begin{figure}
    \centering
    \includegraphics[scale=.35]{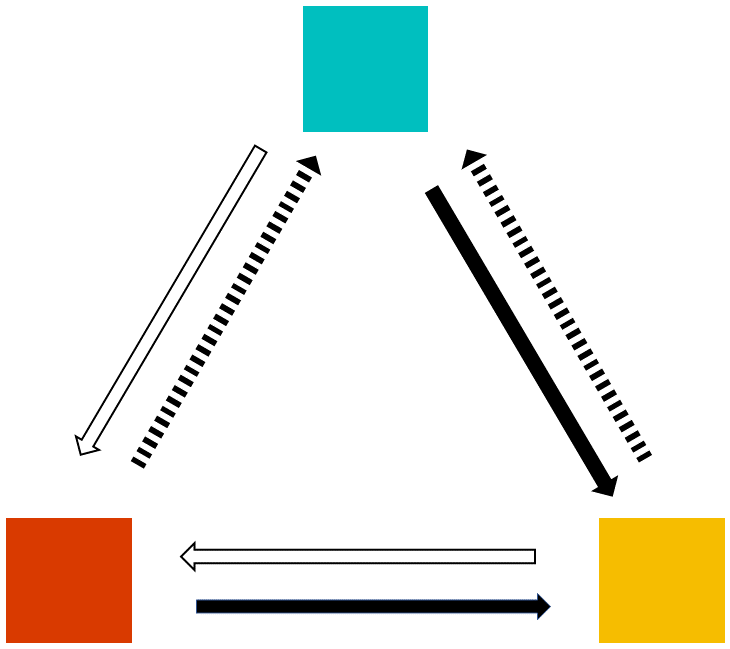}
    \caption{Diagram of different state conversions leading to transportation of singularities along a given contour. Arrows directed between two states represent conversion while arrow color represents contour the singularity will move along.}
    \label{fig:key}
\end{figure}

Although defibrillation is traditionally implemented by excitation of tissue (primarily due to the asymmetric response of virtual electrodes) rearrangement and annihilation of singularities has also been shown to be possible by deexciting cells \cite{keener, ME}. In fact, because external stimulus of the GH model is independent of the rules of regular time evolution, manipulation of singularities is possible by converting any cell state to any other. The result of each conversion is displayed in Figure \ref{fig:key}. For example, converting a singularity-adjacent rest cell to a refractory cell will transport the singularity along the nearby excited front contour. For a given fibrillating configuration such as that of Figure \ref{fig:contoursfib}, a truly optimal defibrillating stimulus could be designed by finding the smallest number of cells necessary to convert in order to annihilate every singularity. This essentially amounts to finding the shortest contour between each pair of singularities and applying the proper conversion from Figure \ref{fig:key} in order to mutually annihilate them. Because singularities are almost never paired exclusively to one other singularity, this problem becomes quite complex as the number of singularities increases.  

Having described how an exciting stimulus interacts with the topological refractory back contours to either increase or decrease the number of spiral core singularities, we now show how these properties can be used to design an optimized pacing scheme. The goal of a low-energy defibrillating shock is to minimize the number of new singularities created while maximizing the number that are eliminated. By monitoring the refractory back contours, shocks can be applied at specific times to achieve this goal.

Our proposed pacing strategy is motivated by the singularity creation mechanism present in e.g. Figure \ref{fig:teleport}(d). In order to completely defibrillate, the refractory back contours joining singularities must be stimulated continuously. Any discontinuities due to the random profile of LEAP shocking will typically generate new pairs of singularities. In fact, the probability of successfully annihilating two singularities joined by a contour of length $L$ is $p^L$ which decreases exponentially with increasing $L$. In light of this fact, a topologically motivated strategy is to shock whenever the total length of the refractory back contours reaches a local minimum. This condition ensures that pairs of singularities are close together and can be annihilated by exciting only a few consecutive cells. Additionally, transient refractory cells from previous shocks contribute significantly to the total contour length. A local minimum in the total length thus indicates that the transients have fully dissipated, and so a new shock will not create new singularities. Since the tracking of contours has been successfully demonstrated in both detailed simulation \cite{robust} and experiment \cite{robustexp} using only noisy voltage data, this defibrillation method is also universally applicable; this reflects the extremely general nature of the topological analysis. 

For comparison, we also demonstrate a second pacing strategy that exploits the Markov formulation of the shock step in terms of the core, vulnerable, and invulnerable patterns. If $C$, $V$, $I$ are the numbers of each type of pattern in a configuration, then the average change in cores after a shock is given from Equation \ref{eq:transition} by
\begin{equation}
    \Delta C = -pC + 2p(1-p)V.
\end{equation}
A simple greedy defibrillation algorithm then consists of computing $\Delta C$ and shocking when it is negative, i.e. when the shock will, on average, decrease the total number of cores. While effective, this method is limited to the GH model for which an explicit Markov representation exists. However, it conveniently constitutes another topologically informed pacing method that can be compared to fixed-period pacing.

\begin{figure}
    \centering
    \includegraphics[scale=.45]{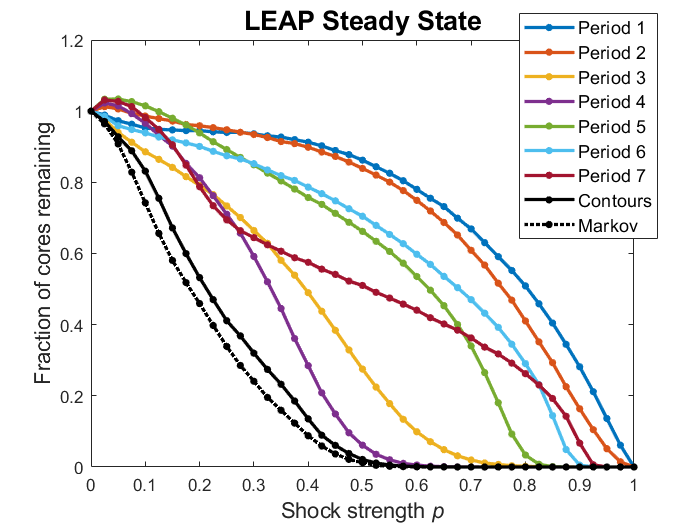}
    \caption{Steady state number of cores in LEAP simulations including Markov and contour pacing strategies.}
    \label{fig:leapfull}
\end{figure}

\begin{figure}
    \centering
    \includegraphics[scale=.45]{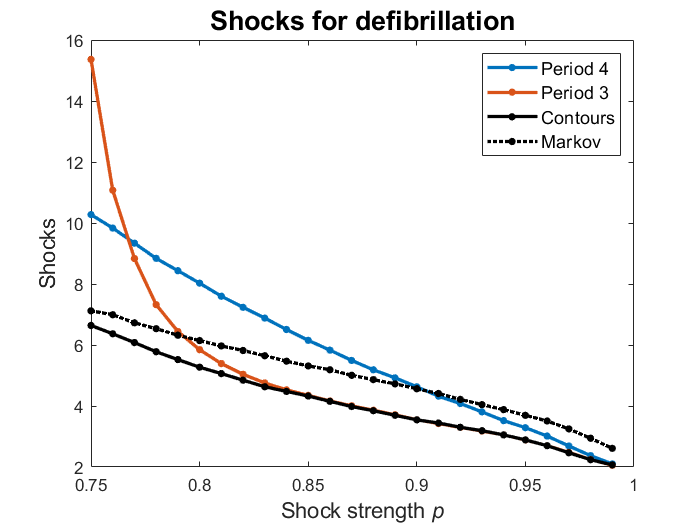}
    \caption{Average number of shocks to reach defibrillation for the four most effective pacing protocols.}
    \label{fig:mfpts}
\end{figure}

\begin{figure}
    \centering
    \includegraphics[scale=.45]{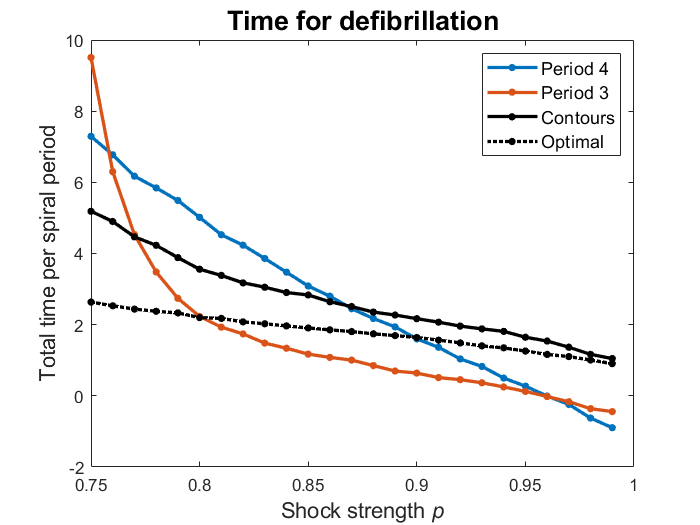}
    \caption{Average time in units of spiral period (four time steps) to reach defibrillation for the four most effective pacing protocols.}
    \label{fig:mfptt}
\end{figure}

Figure \ref{fig:leapfull} shows the steady state fraction of cores for both the Markov and contour methods compared to the original constant period LEAP. Both methods outperform spiral period pacing and successfully defibrillate at a lower shock strength. 
Figures \ref{fig:mfpts} and \ref{fig:mfptt} show the average number of shocks and amount of time, respectively, for the four most effective methods to completely defibrillate. The Markov method typically requires more shocks but consistently defibrillates in the shortest time. The contour method is as effective as period pacing 3 for stronger shock strengths but significantly outperforms it at lower strengths. 

\begin{figure}
    \centering
    \includegraphics[scale=.45]{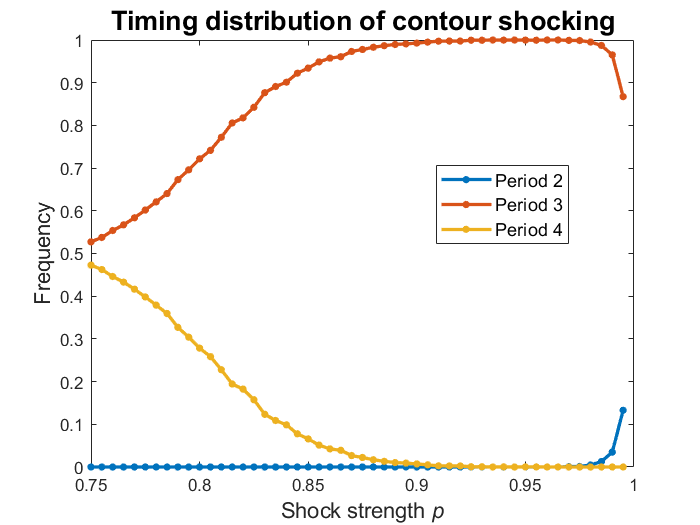}
    \caption{Relative frequency of number of time steps between shocks during the contour tracking protocol.}
    \label{fig:timing}
\end{figure}

This latter relationship can be understood from Figure \ref{fig:timing} which shows the relative frequency of time steps between shocks in the contour tracking method. Initially, shocks occur every three or four steps with similar frequency. By mixing the two timings, the method outperforms constant period pacing at either individual period. As shock strength grows and the mean dissipation time decreases, shocking every three steps becomes more favorable. As shock strength increases further still and the mean dissipation time approaches the minimum of two steps, three-step shocks begin to be replaced by two-step shocks. This progression also demonstrates the contour tracking method's greatest strength: the ability to adapt to different system parameters automatically. Unlike traditional LEAP, which needs to be manually optimized to a particular and non-obvious pacing period which varies with field strength\cite{markusbar}, contour tracking is system-independent and can pick out optimal shock timing on the fly from topological information.

\section{\label{sec:conc}Conclusion}
In this work, we have explored the generic phenomenology of far-field low-energy defibrillation using a cellular automaton model of excitable media. Despite its extreme simplicity, this model obeys the same crucial topological rules as more sophisticated continuum models; fibrillation is produced by the persistence of phase singularities at the cores of spiral waves and ceases only when every pair of singularities is mutually annihilated. From simulations of periodic low-energy pacing, the origin of an optimal pacing period was identified as arising from the competition between two spiral creation mechanisms---one occurring at short period pacing, and one for long. Pacing slowly allows existing spirals to propagate and produce large vulnerable regions where new singularities may be created. Pacing faster than the tissue can return to rest from the previous shock results in many new singularities created along the transient refractory regions. Optimal pacing is achieved by pacing as soon as the refractory transients have dissipated.

We then demonstrated a novel pacing scheme based on established topological analysis of spiral wave turbulence. By tracking the topological contours linking singularities and shocking when their length is short, spiral creation is minimized directly and the resulting shock timing is consistently optimized. Our simulations suggest that this method outperforms traditional fixed-period pacing and defibrillates at a lower shock strength. Most significantly, the topological nature of this strategy makes implementation in clinical experiment feasible; successful tracking of contours from experimental voltage data has been demonstrated in previous studies.

\begin{acknowledgments}
N.D. would like to thank Kurt Wiesenfeld for insightful discussions and guidance.
\end{acknowledgments}

% \nocite{*}
\bibliography{references}% Produces the bibliography via BibTeX.

\end{document}